\documentclass[10pt,conference]{IEEEtran}
\IEEEoverridecommandlockouts
\usepackage{amsmath,amssymb,amsfonts}
\usepackage{algorithmic}
\usepackage{graphicx}
\usepackage{multirow}
\usepackage{float}
\usepackage{latexsym}
\usepackage{array,arydshln}
\usepackage{bm}
\usepackage[outdir=./]{epstopdf}
\usepackage{tikz}
\usepackage{pgfplots}
\usepackage{ctable}

\newcommand{\bs}[1]{\ensuremath{\boldsymbol{#1}}}

\begin{document}

\title{Thresholds of Braided Convolutional Codes \\ on the AWGN Channel}

\author{\IEEEauthorblockN{Muhammad Umar Farooq, Saeedeh Moloudi, and Michael Lentmaier}
\IEEEauthorblockA{Dept. of Electrical and Information Technology, \\
Lund University, Sweden \\
Emails: \{muhammad.umar\_farooq, saeedeh.moloudi, michael.lentmaier\}@eit.lth.se}}



\maketitle


%

\begin{abstract}
In this paper, we perform a threshold analysis of braided convolutional codes (BCCs) on the additive white Gaussian noise (AWGN) channel. The decoding thresholds are estimated by Monte-Carlo density evolution (MC-DE) techniques and compared with approximate thresholds from an erasure channel prediction. The results show that, with spatial coupling, the predicted thresholds are very accurate and quickly approach capacity if the coupling memory is increased. For uncoupled ensembles with random puncturing, the prediction can be improved with help of the AWGN threshold of the unpunctured ensemble.

\end{abstract}

\IEEEpeerreviewmaketitle


\section{Introduction}



Braided convolutional codes (BCCs) \cite{BCCorig} are a class of spatially-coupled (SC) turbo-like codes with regular graph structure. On the binary erasure channel (BEC), explicit density evolution (DE) equations have been derived for BCCs in \cite{DE1_BER}, which can be used to efficiently compute exact decoding thresholds for that channel. The results show that BCCs have superior maximum-a-posteriori (MAP) decoding thresholds compared to parallel or serially concatenated codes on the binary erasure channel (BEC) \cite{SC_TC_DE}.  Furthermore it has been proven analytically in \cite{SC_TC_DE} that threshold saturation occurs, i.e.,  with spatial coupling a belief propagation (BP) decoder can achieve the same threshold as an optimal MAP decoder. 



The aim of this paper is to compute the BP thresholds of BCCs on the additive white Gaussian noise (AWGN) channel. 
For this channel, exact DE equations are not available  for turbo-like codes, and Monte Carlo (MC) methods are usually applied to estimate decoding thresholds. One of the difficulties of such an approach is that the graphs of spatially coupled systems contain a large number of edge types whose message densities have to be considered individually during DE. This requires significantly larger computational efforts than classical methods, like the single edge-type extrinsic information transfer characteristics (EXIT) chart analysis \cite{EXIT}.


We use Monte-Carlo density evolution (MC-DE) to estimate the thresholds of uncoupled and coupled BCCs with and without puncturing.  As an efficient alternative, we then consider the erasure channel approximation by Chung \cite{erasurepred} for predicting the AWGN channel thresholds from those of the BEC and compare the results. Finally, we demonstrate that for randomly punctured ensembles, analogously to LDPC codes \cite{LDPC}, the thresholds of BCCs of all higher rates can immediately be predicted from the unpunctured thresholds of the BEC and/or the AWGN channel. Some simulation results are also given.




\section{Braided Convolutional Codes}

\begin{figure}[tbp]
	\centerline{\includegraphics[scale=.44]{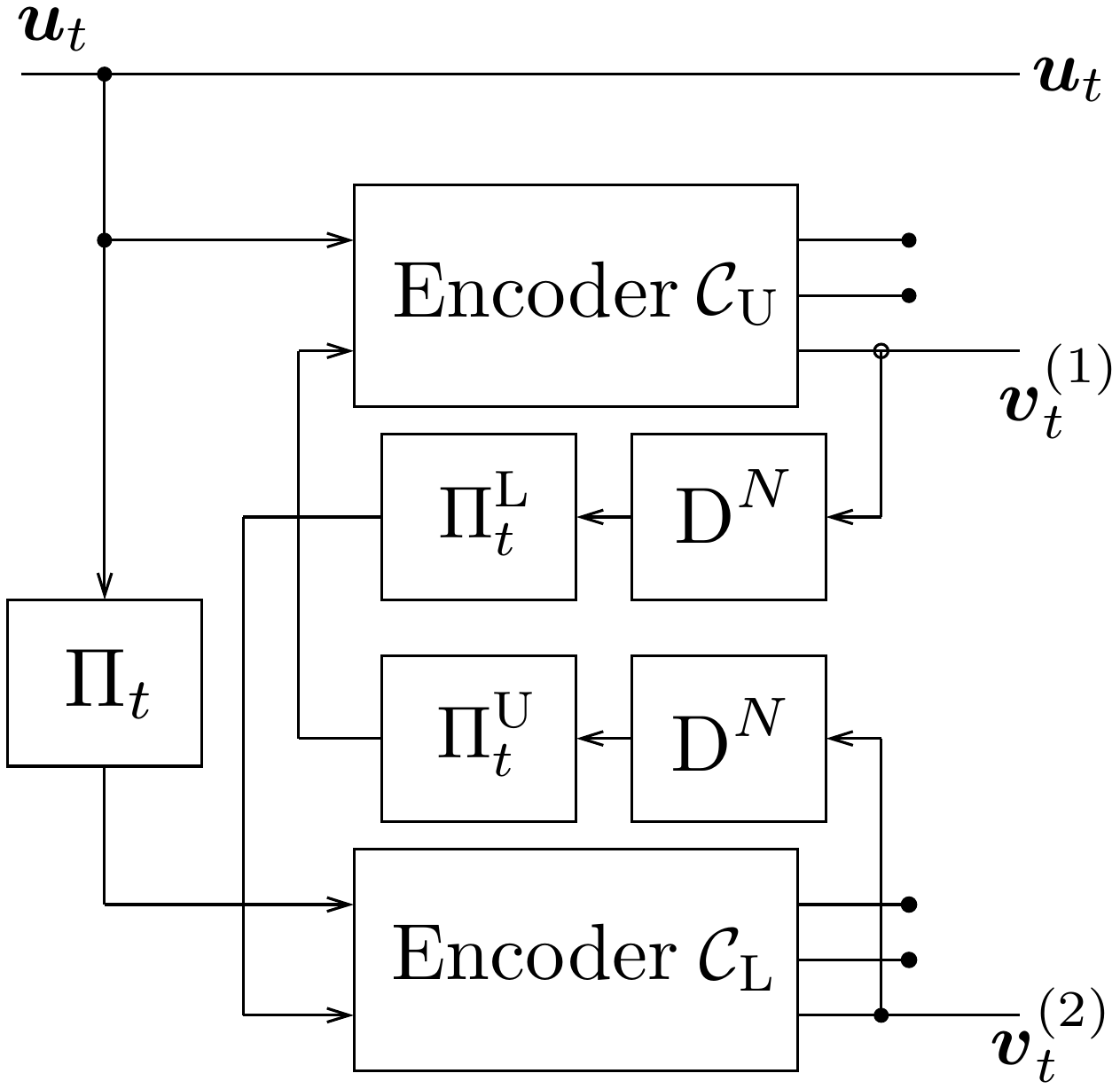}}
	\caption{Blockwise BCCs: turbo-like codes with parity feedback (${R=1/3}$).}
	\label{fig:BCCencoder}
\end{figure}

BCCs were originally introduced in \cite{BCC}. Their characteristics is that the parity symbols of one component encoder are used as information symbols of the other and vice versa. Due to this, both information and parity symbols are protected by both component codes in a symmetric way. In this paper we consider an example of braided convolutional codes (BCCs) of rate $R=1/3$, which are defined by using two systematic component convolutional encoders of rate $R_{cc} =2/3$ and three permutors of length $N$. The component code and the encoding method is same as used in \cite{SC_TC_DE}. In particular, we are using the blockwise BCCs, for which an encoder diagram is shown in Fig. \ref{fig:BCCencoder}. The parity symbols created by one encoder at time $t$ pass a delay of one block, $D^N$ and a permutor before entering the other encoder at time $t+1$ \cite{BCC}.   
 
  
To compare BCCs with PCCs and SCCs, it sometimes can be useful to define an uncoupled equivalent of BCCs. This can be achieved by removing the delay in the encoder depicted in Fig. \ref{fig:BCCencoder}. Since the feedback now occurs without the delay, a straightforward encoding by means of the trellis is not possible. But the code is still well defined by the trellis constraints that the code symbols have to satisfy.




\begin{figure}[htbp]
	\centerline{\includegraphics[scale=.35]{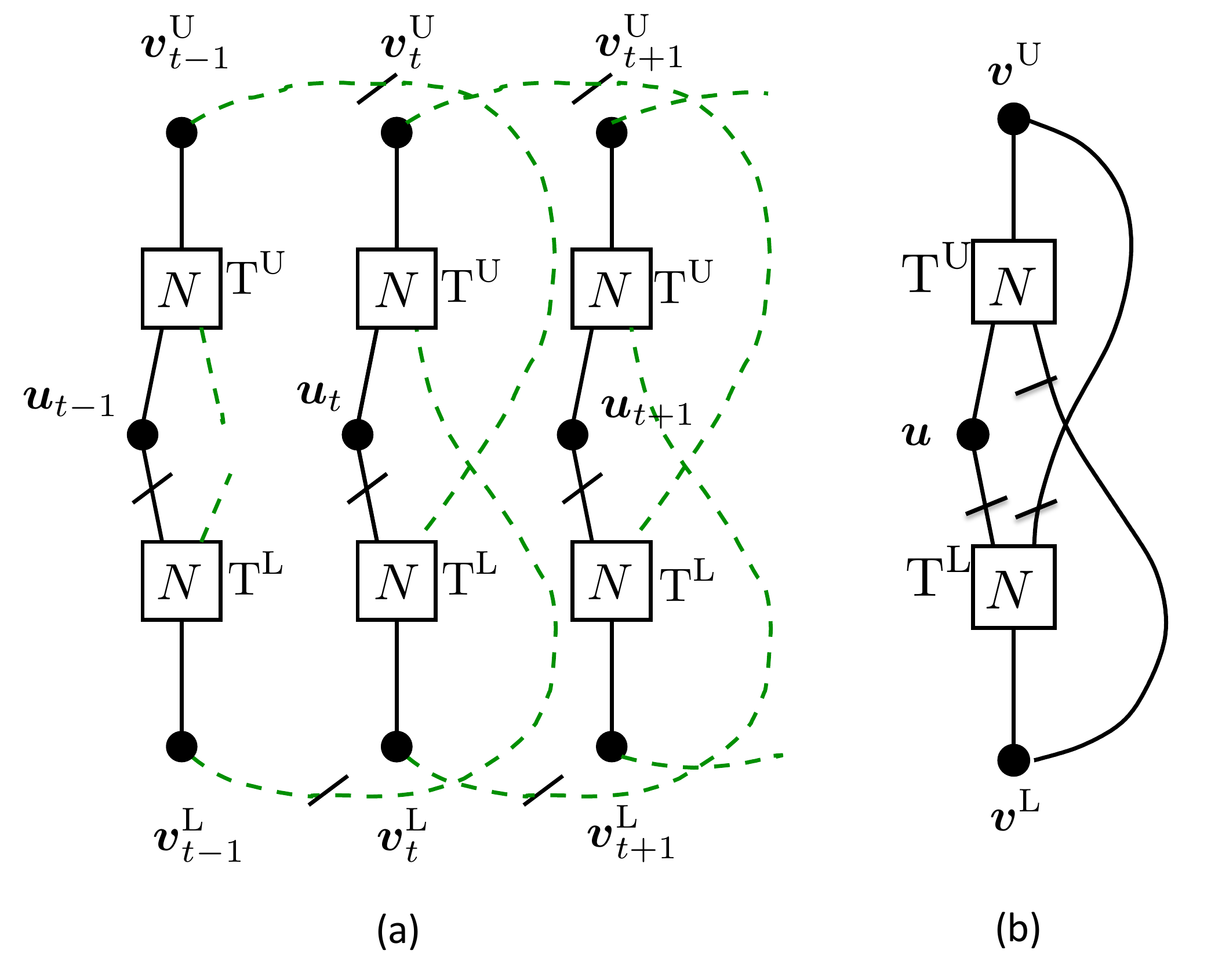}}
	\caption{Compact graph representation }
	\label{fig:compactbcc}
\end{figure}



Turbo-like codes, like LDPC codes, can be described by the factor graphs. This way of expressing the code is useful in describing the exchange of messages in an iterative  BP decoder, as well as for the corresponding DE analysis. Instead of a conventional factor graph, we use a compact graph representation as introduced  in \cite{SC_TC_DE}. The compact graph of a BCC and its uncoupled equivalent is shown in Fig. \ref{fig:compactbcc}. Each block of symbols is represented by a variable node and each trellis by a factor node. A permutor is indicated by a short line that crosses an edge in the graph.

Similar to the SC-LDPC codes, a BCC code can be obtained by repeating the graph of an uncoupled code and spreading some edges across $m+1$ neighboring blocks, where $m$ denotes the coupling memory. The original BCCs shown in Fig. \ref{fig:compactbcc}(a) have coupling memory $m=1$.


\section{Monte Carlo Density Evolution}

\begin{figure}[tbp]
	\centering
	
	\begin{tikzpicture} [thick, scale=0.4]
	\node at (2.5,10.5) {$\bs{L}_\text{ch}^{(1)}$};
	\node at (0.5,9) {$\bs{L}_\text{ext,in}^{(1)}$};
	\node at (3.7,7.6) {$\bs{L}_\text{in}^{(2)}$};
	
	\node at (1,-.7) {$\bs{L}_\text{ch}^{(0)}$};
	\node at (-1,1) {$\bs{L}_\text{ext,in}^{(0)}$};
	\node at (0.3,2.6) {$\bs{L}_\text{in}^{(0)}$};
	
	\node at (4.5,-.7) {$\bs{L}_\text{ch}^{(2)}$};
	\node at (6.3,1) {$\bs{L}_\text{ext,in}^{(2)}$};
	\node at (5.0,2.6) {$\bs{L}_\text{in}^{(1)}$};
	
	\node at (10.5,10.5) {$\bs{L}_\text{ch}^{(1)}$};
	\node at (12.2,7.6) {$\bs{L}_\text{ext,out}^{(2)}$};
	
	\node at (9,-.7) {$\bs{L}_\text{ch}^{(0)}$};
	\node at (7.8,2.8) {$\bs{L}_\text{ext,out}^{(0)}$};

	\node at (12.5,-.7) {$\bs{L}_\text{ch}^{(2)}$};
	\node at (13.1,2.8) {$\bs{L}_\text{ext,out}^{(1)}$};
	
	\draw [thick, dashed][->] (2.5,10) -- (2.5,9.3);
	\draw [thick][->] (2.8,8) -- (2.8,7);
	\draw [thick][->] (1,2) -- (1.35,3);
	\draw [thick][->] (4,2) -- (3.65,3);
	\draw [thick, dashed][->] (1.5,9) -- (2.3,9);
	
	\draw [ultra thick] (1,4) rectangle (4,6);
	\node at (2.5,5.4) {Upper};
	\node at (2.5,4.6) {Trellis};
	\draw [ultra thick, fill] (2.5,9) circle [radius=0.15];
	\draw [ultra thick, fill] (1,1) circle [radius=0.15];
	\draw [thick, dashed][->] (4,0) -- (4,0.8);
	\draw [thick, dashed][->] (0,1) -- (0.8,1);
	\draw [ultra thick, fill] (4,1) circle [radius=0.15];
	\draw [thick, dashed][->] (1,0) -- (1,0.8);
	\draw [thick, dashed][->] (5,1) -- (4.2,1);
	\draw [ultra thick](1,1) -- (2,4);
	\draw [ultra thick](4,1) -- (3,4);
	\draw [ultra thick](2.5,9) -- (2.5,6);
	
	\node at (2.5,-1.5) {(a)};
	
	\draw [thick, dashed][->] (10.5,10) -- (10.5,9.3);
	\draw [thick][<-] (10.8,8) -- (10.8,7);
	\draw [thick][<-] (9,2) -- (9.35,3);
	\draw [thick][<-] (12,2) -- (11.65,3);
	
	\draw [ultra thick] (9,4) rectangle (12,6);
	\node at (10.5,5.4) {Upper};
	\node at (10.5,4.6) {Trellis};
	\draw [ultra thick, fill] (10.5,9) circle [radius=0.15];
	\draw [ultra thick, fill] (9,1) circle [radius=0.15];
	\draw [thick, dashed][->] (12,0) -- (12,0.8);
	\draw [ultra thick, fill] (12,1) circle [radius=0.15];
	\draw [thick, dashed][->] (9,0) -- (9,0.8);
	\draw [ultra thick](9,1) -- (10,4);
	\draw [ultra thick](12,1) -- (11,4);
	\draw [ultra thick](10.5,9) -- (10.5,6);
	
	\node at (10.5,-1.5) {(b)};
	
	\end{tikzpicture}
	\caption{Monte Carlo Density Evolution Process}
	\label{fig:MC-DE}
\end{figure}
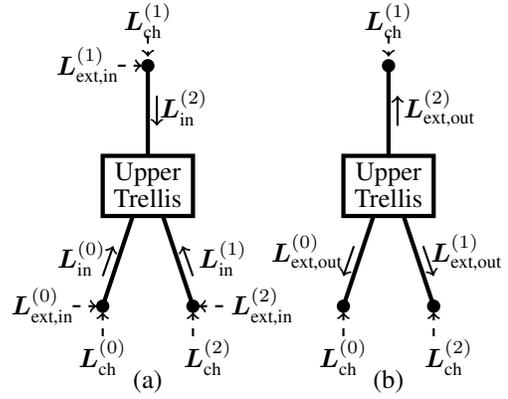

In order to describe the MC-DE process, let us consider the upper decoder of the uncoupled (UC) BCC shown in Fig.~\ref{fig:compactbcc}(b). The exchange of messages with the upper decoder is depicted in Fig.~\ref{fig:MC-DE}. Due to symmetry, the processing at the lower decoder follows analogously. In each iteration of MC-DE, assuming a flooding schedule, the decoding at the upper and lower decoder is performed independently in parallel and the {\em densities} of updated output messages are exchanged. Iterative exchange of densities continues in this fashion until the decoding error probability converges. The key points of the MC-DE process can be summarized in the following three major steps: 

\begin{enumerate}
	\item \textit{Variable node update:} Each variable node $k = 0,1,2$ in Fig.~\ref{fig:MC-DE}(a) takes the sequence  $\bs{L}_\text{ch}^{(k)}$   of channel LLRs and the sequence $\bs{L}_\text{ext,in}^{(j)}$ of incoming extrinsic LLRs  received from output $j$ of the connected lower trellis node and combines them to the updated message sequence $\bs{L}_\text{in}^{(i)} = \bs{L}_\text{ch}^{(k)}+\bs{L}_\text{ext,in}^{(j)}$, which forms the input $i$ of the  upper trellis node. All sequences are of equal length $N$.
	\item \textit{Trellis node update:} The trellis node receives the three input sequences from the different variable nodes corresponding to symbol blocks $\bs{u}$, $\bs{v}^\text{U}$ and $\bs{v}^\text{L}$. The node performs decoding and produces the updated sequences $\bs{L}_\text{ext,out}^{(i)}$ of extrinsic LLRs at each output $i=0,1,2$ of the trellis node. Finally, these output message sequences are used to  estimate the message densities $f({L}_\text{ext,out}^{(i)})$. 
	\item \textit{Drawing samples from message densities:} In this step independent LLR sequences $\bs{L}_\text{ch}^{(k)}$, $k=0,1,2$  and $\bs{L}_\text{ext,in}^{(j)}$, $j=0,1,2$ are created from the channel density and the densities $f({L}_\text{ext,out}^{(j)})$ received from the lower trellis node. These sequences are used in the next MC-DE iteration.  
\end{enumerate}

In the literature,  the message densities $f({L}_\text{ext,out}^{(i)})$ are often {\em approximated} by Gaussian densities as well in MC-DE. In this case, it is not necessary to estimate the exact $f({L}_\text{ext,out}^{(i)})$ in step 2. Instead, some parameter like the mean, the variance or the mutual information corresponding to the LLR sequences is computed to characterize the Gaussian density. 

Table \ref{Table:1UCBCCThd} 
shows the uncoupled BCC thresholds obtained via MC-DE with two different puncturing methods. P2 puncturing refers to when uniform random puncturing is applied on both information and parity bits, whereas in P1  random puncturing is applied only on parity bits. MC refers to the threshold obtained via MC-DE algorithm and GA refers to the threshold when $f({L}_\text{ext,out}^{(i)})$ is approximated to the Gaussian density and $\bs{L}_\text{in}^{(i)}$ are drawn from it. 
\begin{table}[tbp]
	\caption{Uncoupled BCC Thresholds}
	\begin{center}
		\begin{tabular}{lcrrrr}
			\toprule
			\textbf{Thresholds}& \textbf{Pattern} &\multicolumn{3}{c}{\textbf{Rate}} \\
			
			$E_b /N_0$ (dB)& & 1/3 & 1/2 & 2/3 \\    
			\midrule
			
			MC  & P1&1.003 & 2.408 & -\\
			
			GA & P1& 1.023 & 2.399 & -\\
			
			GA SE & P1&1.050 & 2.708 & -\\
			\midrule
			MC & P2&1.003 & 2.121 & 4.151\\
			
			GA & P2&1.018 & 2.128 & 4.062\\
			
			GA SE & P2 &1.048  & 2.161 & 4.131\\
			\bottomrule
			
		\end{tabular}
		\label{Table:1UCBCCThd}
	\end{center}
\end{table}

In determining MC and  GA thresholds, the statistics along the incoming edges to the trellis nodes are different. It means that for a code with multi edge-types (ME), such as BCCs, the $f({L}_\text{in}^{(i)})$ distributions are different along each incoming edge. We can average out the densities $f({L}_\text{in}^{(i)})$ and use a single density $f({L}_\text{in})$ along each incoming edge to the trellis nodes. 

It can be observed that the MC thresholds are closer to the GA thresholds for low rates. Whereas, the difference in these thresholds becomes high as the fraction of puncturing increases. With punctured bits, $f({L}_\text{in}^{(i)})$ will be a mixture of the Gaussian and erasure densities. The erasure density will be dominant in this mixture at higher rate. Therefore we expect that the Gaussian approximation of the densities at higher rate results in inaccuracies. 
Furthermore,  the difference from GA SE thresholds to other thresholds is relatively larger. The GA SE case is equivalent to the classical EXIT chart technique. 
In general, in case of ME ensembles the MC-DE process with GA of densities  $f({L}_\text{in}^{(i)})$ is equivalent to the protograph LDPC EXIT analysis technique \cite{EXITME}.




In order to get a high accurancy of  estimated thresholds via MC-DE, a large number of trellis node output messages are collected before computing the densities within each iteration. The statistics are considered stabilized when the bit error rate (BER) as a function of the number of simulated blocks within an iteration reaches a steady state with a difference of .0001. This requirement of accuracy control makes MC-DE take days to compute UC-BCCs thresholds and it takes even much longer for the SC-BCCs. The accuracy of the MC-DE thresholds can be increased  by running MC-DE for longer time. 

Life can be much more easier if we can predict the thresholds on the Gaussian channel reliably and much faster than the MC-DE does. One way would be to use the exact DE equations as is used in \cite{SC_TC_DE} for BCCs over the BEC. However, the DE equations for the BCCs on the AWGN channel are not available. Therefore, we have to look for an alternative solution to find the thresholds of BCCs much faster than MC-DE does.

\section{Erasure Channel Prediction of AWGN Channel Thresholds}\label{sec:epredmethod}


It has been observed in \cite{erasurepred}  that the thresholds of LDPC codes on the AWGN channel can be {\em approximated} by the  corresponding thresholds on the BEC.  Such an erasure channel prediction (ECP) is computationally  attractive for turbo-like code ensembles, since BEC thresholds can be computed exactly  with relatively small efforts. In this section we will apply this  approach to uncoupled and coupled BCC ensembles and compare the resulting thresholds with those obtained from MC-DE. 



Let $C_E(\varepsilon)=1-\varepsilon$ denote the capacity of a BEC with erasure probability $\varepsilon$ and $C_G(\sigma)$ denote the capacity of a binary-input AWGN channel with noise variance $\sigma^2$. Let $\varepsilon^*$ and $\sigma^*$ denote the DE thresholds computed on the two types of channels for a given code ensemble. The ECP is based on the observation that $C_E(\varepsilon^*) \approx C_G(\sigma^*)$, which suggests the approximation
\begin{equation}\label{eq:mapping}
\sigma^* \approx C_G^{-1}\left(C_E(\varepsilon^*) \right)= C_G^{-1}\left(1-\varepsilon^* \right) \ .
\end{equation}

%

%
%


Using \eqref{eq:mapping}, the AWGN threshold of a given ensemble can now be predicted from the corresponding BEC threshold  $\varepsilon^*$.
Table~\ref{Table:1EbNoBCCThd} shows the resulting predicted thresholds for UC-BCCs and SC-BCCs along with the corresponding MC-DE thresholds. The original BEC thresholds are also given, where the values of puncturing pattern P1 are identical to Table II of \cite{SC_TC_DE}. 
\begin{table}[tbp]
	\caption{Predicted vs Estimated Thresholds ($E_b/N_0$) of BCC }
\begin{center}
\begin{tabular}{lccrrrrr}
	\toprule
	&&\multicolumn{2}{c}{\textbf{Erasure}}& \multicolumn{2}{c}{\textbf{UC BCC}} &\multicolumn{2}{c}{\textbf{SC BCC }} \\
	
	Rate & Pattern &$\epsilon_{SC}^1$ & $\epsilon_{UC}$ & ECP & MC & ECP&  MC\\    
	
	\midrule
	
	1/3 &P1&0.6609& 0.5541& 1.213 & 1.00 &-0.399 & -0.39
	
	\\
	
	1/2 &P1&0.4932&0.3013& 2.716 & 2.40 &0.276& 0.25
	
	\\
	
	2/3 &P1&0.3257& -&-&-&1.156&1.12
	
	\\
	
	3/4 &P1&0.2411& -&-&-&1.746&1.70
	
	\\
	
	4/5 &P1&0.1915&- &-&-&2.164&2.12 \\
	
\midrule
	
	1/3 & P2 & 0.6609&0.5541&1.213&1.00&-0.400&-0.39
	
	\\
	
	1/2 & P2 &0.4914&0.3312&2.335&2.12&0.300&0.29
	
	\\
	
	2/3 & P2&0.3219&0.1083&4.336&4.15&1.204&1.18
	
	\\
	
	3/4 & P2&0.2371&-&-&-&1.800&1.77
	
	\\
	
	4/5 & P2 &0.1862 &-&- &-&2.241&2.21 \\
	
	\bottomrule
	
\end{tabular}
\label{Table:1EbNoBCCThd}
\end{center}
\end{table}
%
We can see from Table~\ref{Table:1EbNoBCCThd} that the ECP thresholds of SC-BCCs are quite close to the MC thresholds. However, the difference is larger on higher rates than it is on lower rates as we have observed in the UC-BCCs thresholds. The ECP and MC thresholds of the UC-BCCs have bigger difference than the SC-BCCs. Furthermore, the ECP and the MC thresholds are much closer to each other on P2 puncturing compared to  P1 puncturing. 





\section{ Thresholds of Randomly Punctured Ensembles}\label{sec:thdPred}

Can we predict the thresholds such that the difference of the predicted and the estimated threshold do not increase as the fraction of puncturing increases? In \cite{LDPC}, it has been shown that for the P2 punctured LDPC codes, it is possible to closely predict the thresholds on the AWGN channel, given just the threshold of the unpunctured code ensemble  and the design rate $R$. In the following, we will apply the methods discussed in \cite{LDPC} to the randomly punctured BCC ensembles and discuss the results obtained by it.

\subsection{$\theta_E$ Predictions }

This method takes only the unpunctured code threshold on the BEC and predicts the thresholds for the punctured ensembles on the BEC by using 
\begin{equation}\label{eq:erasure}
 \epsilon^*(\alpha) = 1 - \theta_E R(\alpha ) \ , 
\end{equation}
where
\begin{equation}\label{eq:erasureslope}
\theta_E = \frac{1-\epsilon^*}{R} \ .
\end{equation}
The elegance of this method is that it will provide the exact BEC thresholds for all  rates $R(\alpha)$, $R \leq R(\alpha )  \leq 1/\theta_E$, based on a single parameter $\theta_E$ and hence it is not required to perform DE on all different rates. However, this method is limited to the puncturing pattern P2 only.

Once we know the BEC thresholds, we can apply the ECP method discussed in Section \ref{sec:epredmethod}. Equivalently, we can write
\begin{equation}\label{eq:erasure}
h_G(\sigma^*(\alpha)) \approx  h_E(\varepsilon^*(\alpha)) = \epsilon^*(\alpha) = 1 - \theta_E R(\alpha ) \ , 
\end{equation}
where $h_G(\sigma)=1-C_G(\sigma)$ and $h_E(\varepsilon)=1-C_E(\varepsilon)=\varepsilon$ denote the conditional entropy of the AWGN channel and the BEC, respectively. From \eqref{eq:erasure} we can compute the AWGN thresholds in terms of standard deviation $\sigma$ or signal-to-noise ratio $E_b/N_0$.
The results obtained via this method for UC-BCCs are given in Table \ref{Table:UCBCCThd}.

\begin{table}[tbp]
	\caption{Uncoupled BCC Thresholds}
	\begin{center}
		\begin{tabular}{lcrrrr}
			\toprule
			\textbf{Thresholds}& \textbf{Pattern} &\multicolumn{3}{c}{\textbf{Rate}} \\
			
			$E_b /N_0$ (dB)& & 1/3 & 1/2 & 2/3 \\    
			\midrule
			
			Erasure Probability &P2& 0.5541 & 0.3312 & 0.1083 \\
			
			MC-DE &P2& 1.003 & 2.121 & 4.151  \\
			
			$\theta_E$ Predicted &P2 & 1.214 & 2.336 & 4.338 \\
			
			$\theta_G$ Predicted &P2&1.003 & 2.054 & 3.798\\
			
			MP & P2& 1.003 & 2.195 & 4.197  \\

			\bottomrule
			
		\end{tabular}
		\label{Table:UCBCCThd}
	\end{center}
\end{table}

\subsection{$\theta_G$ Predictions}
While the $\theta_E$ prediction works especially well for high code rates, it can be further improved for low rates if the unpunctured threshold $\sigma^*$ is available. Then
\begin{equation} \label{eq:entropygauss} 
h_G(\sigma^*(\alpha)) \approx 1 - \theta_G R(\alpha ) \ ,
\end{equation}
where
\begin{equation}\label{eq:gaussian}
\theta_G = \frac{1-h_G(\sigma^*)}{R} \ .
\end{equation}

It can be seen in Table \ref{Table:UCBCCThd} that for the lower code rates, the MC thresholds and the $\theta_G$ predictions are quite close. However, for a higher fraction of punctured bits, the MC thresholds are closer to the $\theta_E$ predictions. 

\subsection{Mixed Predictions}

To account for the higher puncturing fraction in the prediction of thresholds on the AWGN channel, a mixed prediction (MP) method is suggested in \cite{LDPC}, where both $\theta_E$ and $h_G(\sigma^*)$ are used. 

The important idea behind the mixed prediction method is that various punctured code ensemble thresholds on the BEC, when viewed in the entropy domain, lie on a straight line. The slope of this line is characterized by $\theta_E$. It is further demonstrated in \cite{LDPC} that for P2 punctured LDPC codes on the AWGN channel the estimated entropies, corresponding to the $E_b/N_0$ thresholds, are observed to follow a straight line. 

\begin{figure}[tbp]
	\centerline{\includegraphics[scale=.6]{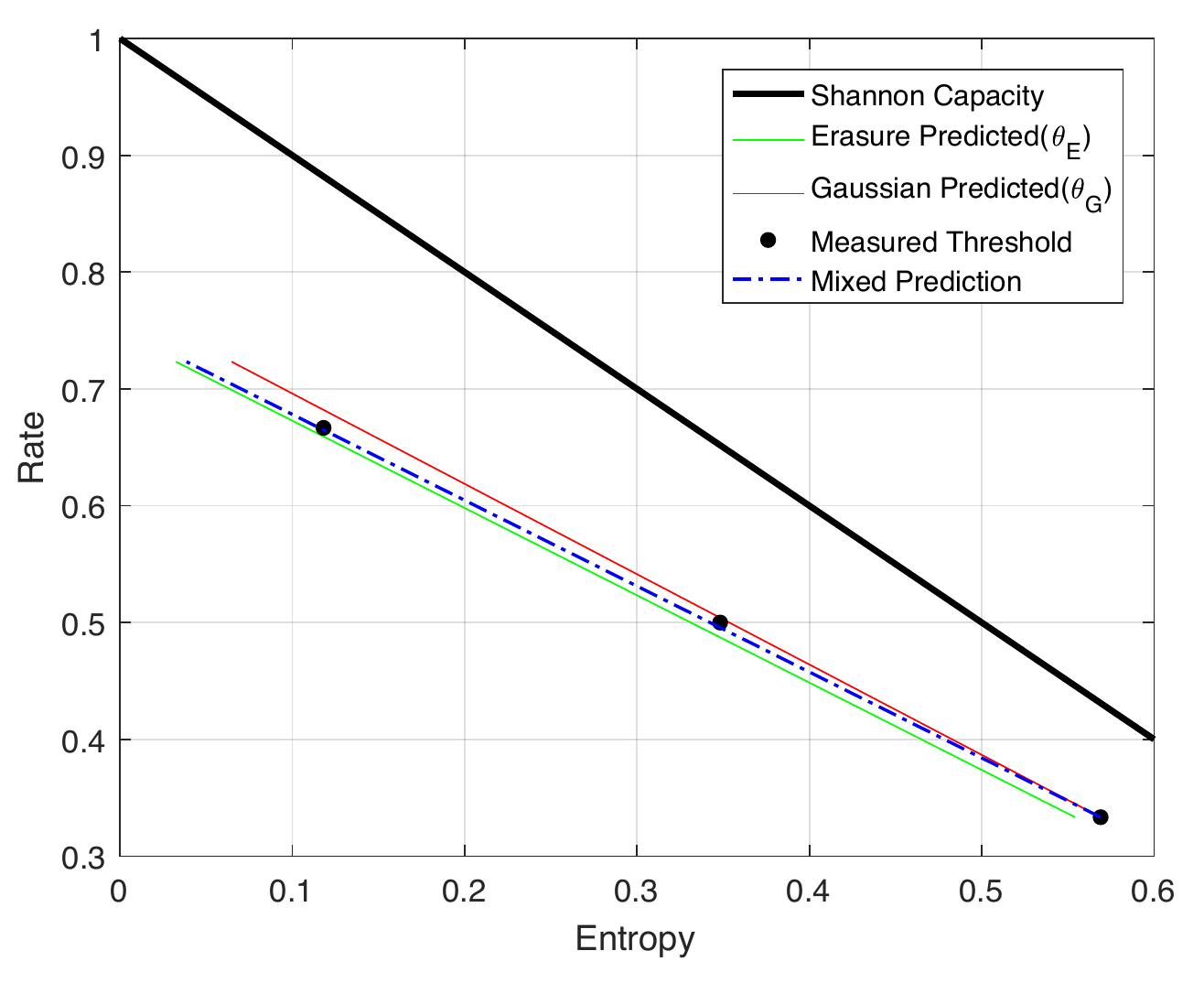}}
	\caption{Entropy vs Rate of UC-BCCs}
	\label{fig:EntropyRateUC}
\end{figure}

From the observation that at lower rate the estimated thresholds are closer to the $\theta_G$ predictions and at higher rates are closer to the $\theta_E$ predictions, we obtain the following mixed prediction
\begin{equation} \label{eq:mixedpred}
\begin{split}
h_G(\sigma_{BP}(\alpha)) \approx 
\frac{\left( R(\alpha)-R \right) \cdot \left( 1-\theta_E-h_G(\sigma^*) \right) }{1-R} & \\ + h_G(\sigma^*)
\end{split}
\end{equation}
where $R \leq R(\alpha )  \leq R_{max}$. $R_{max}$ is an intercept that can be obtained by setting $h_G(\sigma_{BP}(\alpha)) =0$ in \eqref{eq:mixedpred}. This straight line prediction  passes through the unpunctured rate $R(0)$ and the associated threshold (entropy) on the AWGN channel and $R(\alpha)=1$ and its associated threshold (entropy) on the BEC.
 The results of the mixed predictions are shown in Fig. \ref{fig:EntropyRateUC} for the P2 punctured UC-BCCs ensembles. The estimated MC-DE thresholds almost follow the mixed prediction line in this figure. However, the estimated thresholds do not strictly lie on a straight line as was the case in \cite{LDPC}.



\begin{figure}[tbp]
 \centerline{\includegraphics[scale=.6]{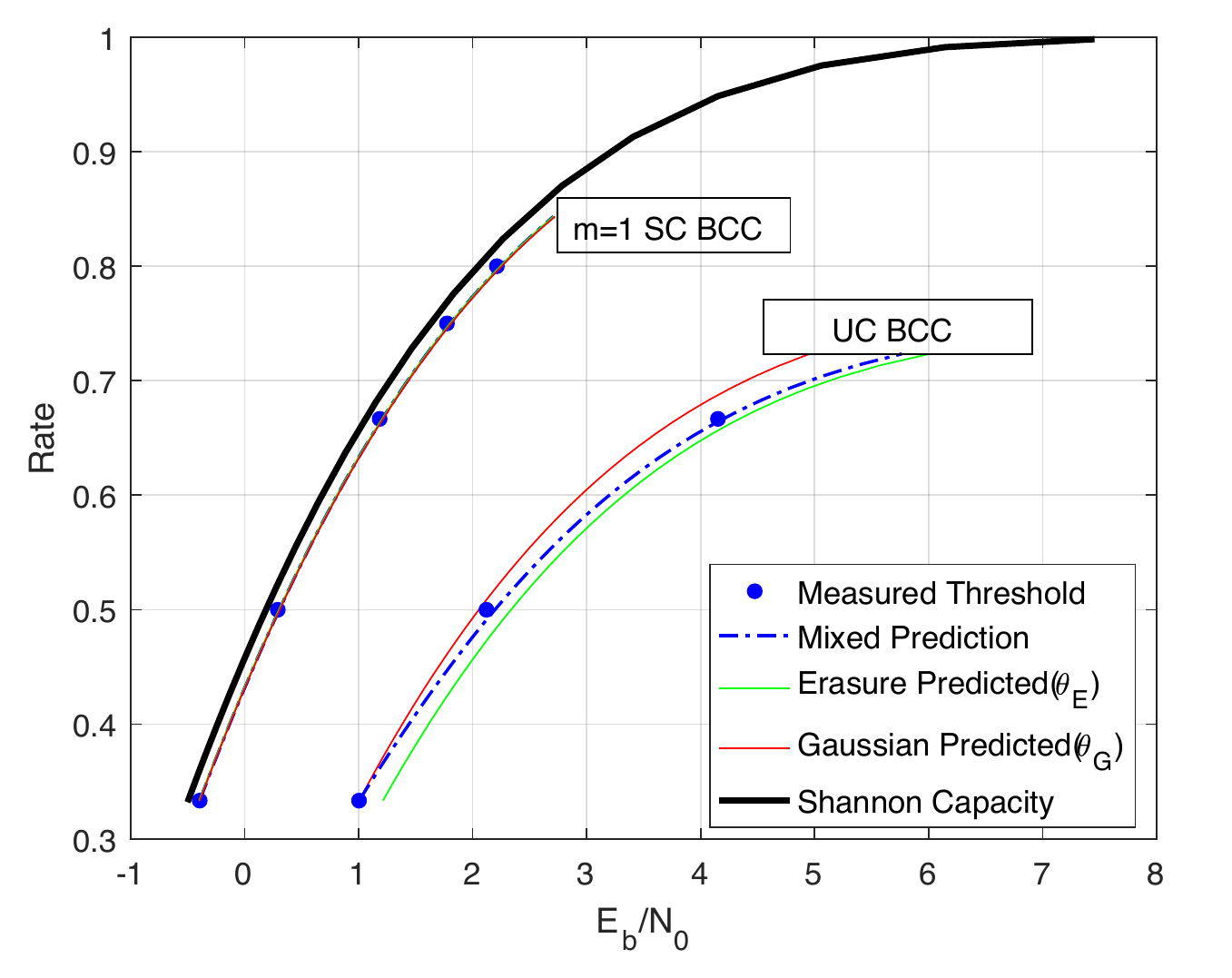}}
 \caption{$E_b/N_0$ vs Rate of SC and UC-BCCs}
 \label{fig:EbNoRateSCTIMIUCBCC}
 \end{figure}

It can be seen in Table \ref{Table:1EbNoBCCThd} that for SC-BCCs, the MC-DE and predicted thresholds are almost identical. 
As a result of threshold saturation, the SC-BCCs performance is much better compared to the performance of UC- BCCs and the gap to capacity is much smaller for SC-BCCs, which can be seen in Fig. \ref{fig:EbNoRateSCTIMIUCBCC}. $\theta_E$ predictions for the punctured $m=1$ SC-BCCs have been made by using $\theta_E = 1.017 $. Similary, $\theta_E$ and $\theta_G$ predictions for the punctured UC-BCCs have been made by using $\theta_E = 1.337 $ and $\theta_G = 1.293 $ respectively. Since the $\theta_E$ predictions are very close to the MC-DE thresholds, $\theta_G$ and mixed predictions have not been included in this table. 




Consider now a larger coupling memory $m=3$. The predicted thresholds are given in Table \ref{Table:CBCCTIMIII}, where it can be seen that $m=3$ SC-BCCs outperform $m=1$ SC-BCCs and operate very close to the Shannon limit. Since, for  $m=3$ SC-BCCs, the $\theta_E$ predictions and MC-DE thresholds are very close to each other, we have only provided the $\theta_E$ predictions. 

\begin{table}[tbp]
	\caption{BCC coupled - $m = 3$}
	\begin{center}
		\begin{tabular}{lcrrrr}
			\toprule
			\textbf{Thresholds} &\multicolumn{5}{c}{\textbf{Rate}} \\
			
			$E_b /N_0$ (dB)& 1/3 & 1/2 & 2/3 & 3/4 & 4/5\\    
			\midrule
			
			Shannon Capacity &   -0.50 & 0.18   & 1.06 &  1.63 &  2.05  \\
			
			Erasure Probability & 0.6644  &  0.4967 &  0.3289  & 0.2450 & 0.1947   \\
			
			$\theta_E$ prediction& -0.4585  &  0.2307  & 1.1151 & 1.6924 & 2.1166  \\

			\bottomrule
			
		\end{tabular}
		\label{Table:CBCCTIMIII}
	\end{center}
\end{table}





\subsection{Simulation Results}

\begin{figure}[tbp]
	\centering
	\includegraphics[scale=.5]{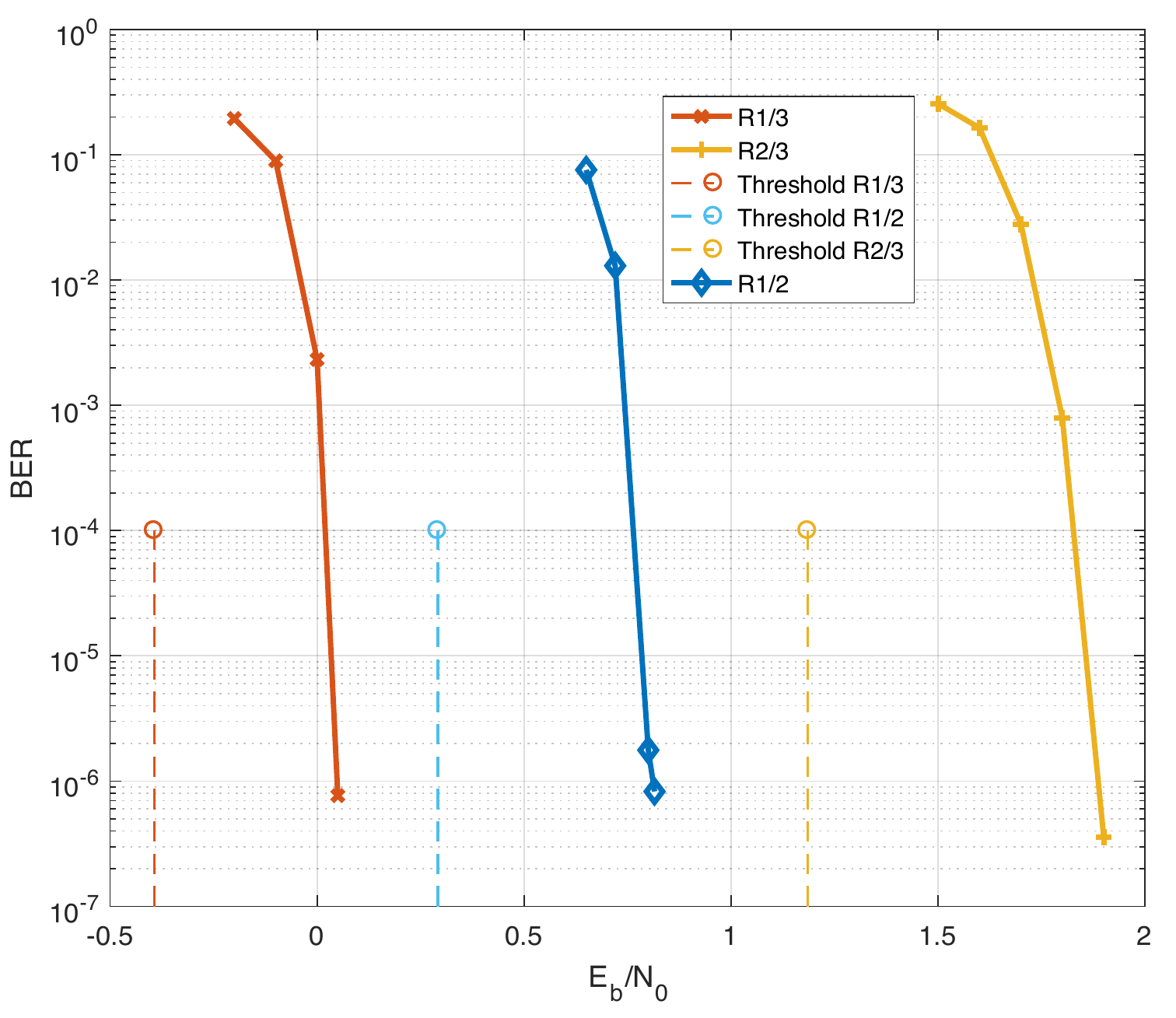}
	\caption{BER vs $E_b/N_0$ thresholds of P2 punctured SC-BCCs}
	\label{fig:EbNoRateSCTIMI}
\end{figure}

The simulated BER performance of P2 punctured $m=1$ SC-BCCs on the AWGN channel  is shown in Fig.~\ref{fig:EbNoRateSCTIMI}.
%
%
The simulations are obtained with a sliding window decoder  \cite{BCCslidingwindow} with window size  $w=5$, and  20 iterations at each window position. 
%
%
%
%
We can see from Fig. \ref{fig:EbNoRateSCTIMI} that  for all the considered rates $R=1/3, 1/2, 2/3$ the BER curves are in accordance with their corresponding thresholds.

\section{Conclusions}

In this paper, have presented MC-DE as a technique to compute the AWGN channel thresholds of spatially-coupled turbo-like codes. Furthermore, we have applied  the threshold prediction methods presented for LDPC codes in  \cite{LDPC} for predicting  BCC thresholds. The $\theta_E$ and $\theta_G$ predictions of the UC and SC-BCC ensembles have been compared with the estimated thresholds obtained by MC-DE. The results show that, with spatial coupling, the predicted thresholds are very accurate and quickly approach capacity if the coupling memory is increased. For uncoupled ensembles with random puncturing, the prediction can be improved with help of the AWGN threshold of the unpunctured ensemble.

%
%




\end{document}